\documentclass{sig-alternate-05-2015}

\usepackage{epsfig,endnotes}
\usepackage{url}
\usepackage{cite}
\usepackage{color}
\usepackage{hyperref}
\usepackage{latexsym}
\usepackage{graphicx}
\usepackage{amsfonts}
\usepackage{amsmath}
\usepackage{footnote}     
\usepackage{float}        
\restylefloat{table}      
\usepackage{wrapfig}      
\usepackage{subcaption}   
\usepackage{tabularx}
\usepackage[font=small]{caption} 
\usepackage[toc,page]{appendix} 
\usepackage{enumitem}     
\usepackage{multirow}

\newcommand{\paragraphbe}[1]{\vspace{0.75ex}\noindent{\bf \em #1} }

\newcolumntype{C}[1]{>{\centering\let\newline\\\arraybackslash\hspace{0pt}}m{#1\textwidth}}
\newcolumntype{L}[1]{>{\centering\let\newline\\\arraybackslash\hspace{0pt}}m{#1\textwidth}}

\begin{document}
\title{Defeating Image Obfuscation with Deep Learning}

\numberofauthors{3}
\author{
\alignauthor Richard McPherson \\
\affaddr{The University of Texas at Austin}\\
\email{richard@cs.utexas.edu}
\alignauthor Reza Shokri \\
\affaddr{Cornell Tech}\\
\email{shokri@cornell.edu }
\alignauthor Vitaly Shmatikov \\
\affaddr{Cornell Tech}\\
\email{shmat@cs.cornell.edu}
}



\maketitle

\begin{abstract}
{ 
We demonstrate that modern image recognition methods based on artificial
neural networks can recover hidden information from images protected by
various forms of obfuscation.  The obfuscation techniques considered in
this paper are mosaicing (also known as pixelation), blurring (as used
by YouTube), and P3, a recently proposed system for privacy-preserving
photo sharing that encrypts the significant JPEG coefficients to make
images unrecognizable by humans.  We empirically show how to train
artificial neural networks to successfully identify faces and recognize
objects and handwritten digits even if the images are protected using
any of the above obfuscation techniques.  } 
\end{abstract}

\section{Introduction}

As user-contributed photographs and videos proliferate online social networks,
video-streaming services, and photo-sharing websites, many of them can be found
to leak sensitive information about the users who upload them, as well as
bystanders accidentally captured in the frame.  In addition to the obvious
identifiers such as human faces, privacy can be breached by images of physical
objects, typed or handwritten text, license plates, contents of computer
screens, etc.

Fully encrypting photos and videos before sharing or uploading them
blocks direct information leakage, but also destroys information that is
not privacy-breaching.   Furthermore, encryption prevents many common
forms of processing.  For example, encrypted photos cannot be easily
compressed for network transmission or cloud storage.

Several privacy protection technologies aim to solve this challenge by
obfuscating or partially encrypting the sensitive parts of the image while
leaving the rest untouched.  A classic example of such a technology is facial
blurring.  It suppresses recognizable facial features but keeps everything else
intact, thus preserving the utility of photographs for purposes such as news
reporting.  These techniques do not provide well-defined privacy guarantees.
Privacy is argued informally, by appealing to the human users' inability to
recognize faces and other sensitive objects in the transformed images.

We argue that humans may no longer be the ``gold standard'' for extracting
information from visual data.  Recent advances in machine learning based on
artificial neural networks have led to dramatic improvements in the state of the
art for automated image recognition.  Trained machine learning models now
outperform humans on tasks such as object
recognition~\cite{he2015delving,ioffe2015batch} and determining the geographic
location of an image~\cite{weyand2016planet}.  In this paper, we investigate
what these advances mean for privacy protection technologies that rely on
obfuscating or partial encrypting sensitive information in images.

\paragraphbe{Our contributions.}  
We empirically demonstrate how modern image recognition techniques based on
artificial neural networks can be used as an adversarial tool to recover hidden
sensitive information from ``privacy-protected'' images.  We focus on three
privacy technologies.  The first is mosaicing (pixelation), which is a popular
way of obfuscating faces and numbers.  The second is face blurring, as deployed
by YouTube~\cite{youtubeface}.  The third is P3~\cite{p3}, a recently proposed
system for privacy-preserving photo sharing that encrypts the significant
coefficients in the JPEG representation of the image.  P3 aims to make the
image unrecognizable yet preserve its JPEG structure and enable servers to
perform certain operations on it (e.g., compress it for storage or
transmission).

To illustrate how neural networks defeat these privacy protection
technologies, we apply them to four datasets that are often used
as benchmarks for face, object, and handwritten-digit recognition.
All of these tasks have significant privacy implications.  For example,
successfully recognizing a face can breach the privacy of the person
appearing in a recorded video.  Recognizing digits can help infer the
contents of written text or license plate numbers.

On the MNIST dataset~\cite{mnist} of black and white hand-written digits, our
neural network achieves recognition accuracy of almost 80\% when the images are
``encrypted'' using P3 with threshold level 20 (a value recommended by the
designers of P3 as striking a good balance between privacy and utility).  When
the images are mosaiced with windows of resolution $8 \times 8$ or smaller,
accuracy exceeds 80\%.  By contrast, the accuracy of random guessing is only
10\%.

On the CIFAR-10 dataset~\cite{cifar} of colored images of vehicles and
animals, we achieve accuracy of 75\% against P3 with threshold 20, 70\%
for mosaicing with $4 \times 4$ windows, and 50\% for mosaicing with $8
\times 8$ windows, vs.\ 10\% for random guessing.

On the AT\&T~\cite{att} dataset of black-and-white faces from 40 individuals, we
achieve accuracy of 57\% against blurring, 97\% against P3 with threshold 20,
and over 95\% against all examined forms of mosaicing, vs.\ 2.5\% for random
guessing.

On the FaceScrub~\cite{facescrub} dataset of photos of over 530
celebrities, we achieve accuracy of 57\% against mosaicing with $16
\times 16$ windows and 40\% against P3 with threshold 20, vs.\ 0.19\%
for random guessing.

The key reason why our attacks work is that we do \emph{not} need
to specify the relevant features in advance.  We do not even need
to understand what exactly is leaked by a partially encrypted or
obfuscated image.  Instead, neural networks automatically discover
the relevant features and learn to exploit the correlations between
hidden and visible information (e.g., significant and ``insignificant''
coefficients in the JPEG representation of an image).  As a consequence,
obfuscating an image so as to make it unrecognizable by a human may no
longer be sufficient for privacy.

In summary, this paper is the first to demonstrate the power of modern neural
networks for adversarial inference, significantly raising the bar for the
designers of privacy technologies.

\section{Image obfuscation}

\begin{figure}[tb]
  \captionsetup[subfigure]{justification=centering}
  \centering
  \includegraphics[width=\columnwidth]{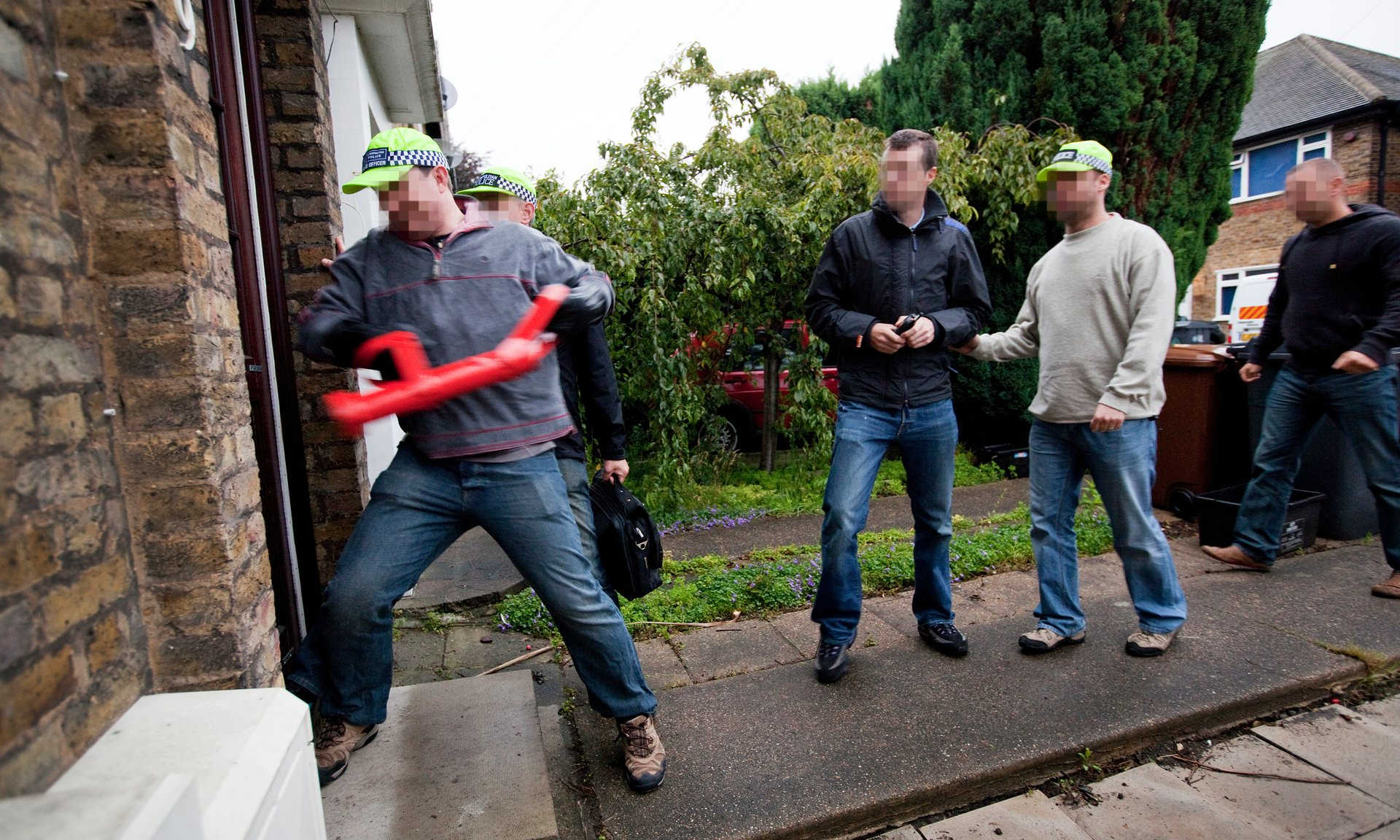}
  \caption{An image from \textit{The Guardian} showing a police raid on a drug
    gang~\cite{warondrugs-pic}. The accompanying article explains that UK drug
    gangs are growing more violent and that police informants and undercover
    operatives face possible retaliation~\cite{warondrugs}. The officers' faces
    are presumably mosaiced for their protection. The window appears to be
    $12\times12$ pixels. Using $16\times16$ windows (which obfuscate more
    information than $12\times12$ windows), our neural network achieves 57\%
    accuracy in recognizing an obfuscated image from a large dataset of 530
    individuals.  The accuracy increases to 72\% when considering the top five
    guesses.}
 \label{fig:war-on-drugs}
\end{figure}

As the targets for our analysis, we chose three obfuscation techniques
that aim to remove sensitive information from images.  The first two
techniques are \emph{mosaicing} (or pixelation) and \emph{blurring}.
These are very popular methods for redacting faces, license plates, adult
content, and text (Figure~\ref{fig:war-on-drugs}).  The third technique
is a recently proposed system called P3~\cite{p3} that aims to protect
the privacy of images uploaded to online social networks such as Facebook
while still enabling some forms of image processing and compression.

\subsection{Mosaicing}

Mosaicing (pixelation) is used to obfuscate parts of an image.  The section to
be obfuscated is divided into a square grid.  We refer to the size of each
square (a.k.a., ``pixel box'') as the mosaic window.  The average color of every
pixel in each square is computed and the entire square is set to that
color~\cite{hill2016effectiveness}.

The size of the window can be varied to yield more or less privacy.  The larger
the box, the more pixels will be averaged together and the less fine-grained the
resulting mosaiced image will be.

Although the size of the image stays the same, mosaicing can be thought
of as reducing the obfuscated section's resolution. For example, a window
of size $n \times n$ applied to an image effectively reduces the number
of unique pixels in an image by a factor of $n^2$.

\subsection{Blurring}

\begin{figure}[tb]
  \captionsetup[subfigure]{justification=centering}
  \centering
  \includegraphics[width=0.4\columnwidth]{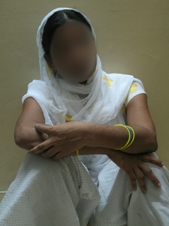}
  \caption{A victim of human trafficking in India~\cite{girlalone-pic}. Her face
    has been blurred, presumably to protect her identity. Our neural networks,
    trained on black-and-white faces blurred with YouTube, can identify a
    blurred face with over 50\% accuracy from a database of 40 faces.}
 \label{fig:blurredgirl}
\end{figure}

Blurring (often called ``Gaussian blur'') is similar to mosaicing and used
to obfuscate faces and sensitive text.  Blurring removes details from an
image by applying a Gaussian kernel~\cite{hill2016effectiveness}. The
result is a ``smoothed'' version of the original image (see
Figure~\ref{fig:blurredgirl}).

In 2012, YouTube introduced the ability to automatically blur all faces
in a video~\cite{youtubeface}. YouTube presents their automatic facial
blurring as a way to improve video privacy.  For example, they suggest
that it can be used it to ``share sensitive protest footage without
exposing the faces of the activists involved.'' In 2016 YouTube introduced
the ability to blur arbitrary objects in a video~\cite{youtubeblur}. Once
this feature is selected, YouTube will attempt to continue blurring the
selected objects as they move around the screen.  YouTube claims to have
added this feature ``with visual anonymity in mind.''

Mosaicing and blurring do not remove all information from an image, but aim to
prevent human users from recognizing the blurred text or face.  The result of
these techniques also often used as it is less visually jarring than, for
example, a black box produced by full redaction.

In Section~\ref{sec:related}, we survey prior papers that demonstrated the
insufficiency of mosaicing and blurring as a privacy protection technique.
To the best of our knowledge, this paper is the first to demonstrate that
standard image recognition models can extract information from mosaiced
and blurred images.

\subsection{P3}

Privacy Preserving Photo Sharing (P3)~\cite{p3} was designed to protect
the privacy of JPEG images hosted on social media sites.  P3 is intended
to be applied by users, but it also assumes that the sites will not
prevent users from uploading P3-protected images.

The main idea of P3 is to split each JPEG into a public image and a
secret image.  The public image contains most of the original image
but is intended to exclude the sensitive information.  Public images
can thus be uploaded to online social networks and untrusted servers.
It is essential that the public image is not encrypted and correctly
formatted as a JPEG since some popular sites (e.g., Facebook) do not
allow users to upload malformed images.

The secret image is much smaller in size but contains most of the
sensitive information from the original image. It is encrypted and
uploaded to a third-party hosting service like Dropbox.

Given a public image and a private image, it is easy to combine them back into
the original image.  P3 proposes a browser plugin that can be used while
browsing social media sites.  It automatically downloads an encrypted image,
decrypts it if the user has the appropriate keys, and displays the combined
image.  Anyone browsing the social media site without the plugin or the
corresponding keys would only see the public parts of images.

P3 supports the recovery of original images even after transformations
(e.g., cropping and resizing) are applied to the public image, but the
details of this do not affect the privacy properties of P3.

\begin{figure}[tb]
  \captionsetup[subfigure]{justification=centering}
  \centering
  \includegraphics[width=\columnwidth]{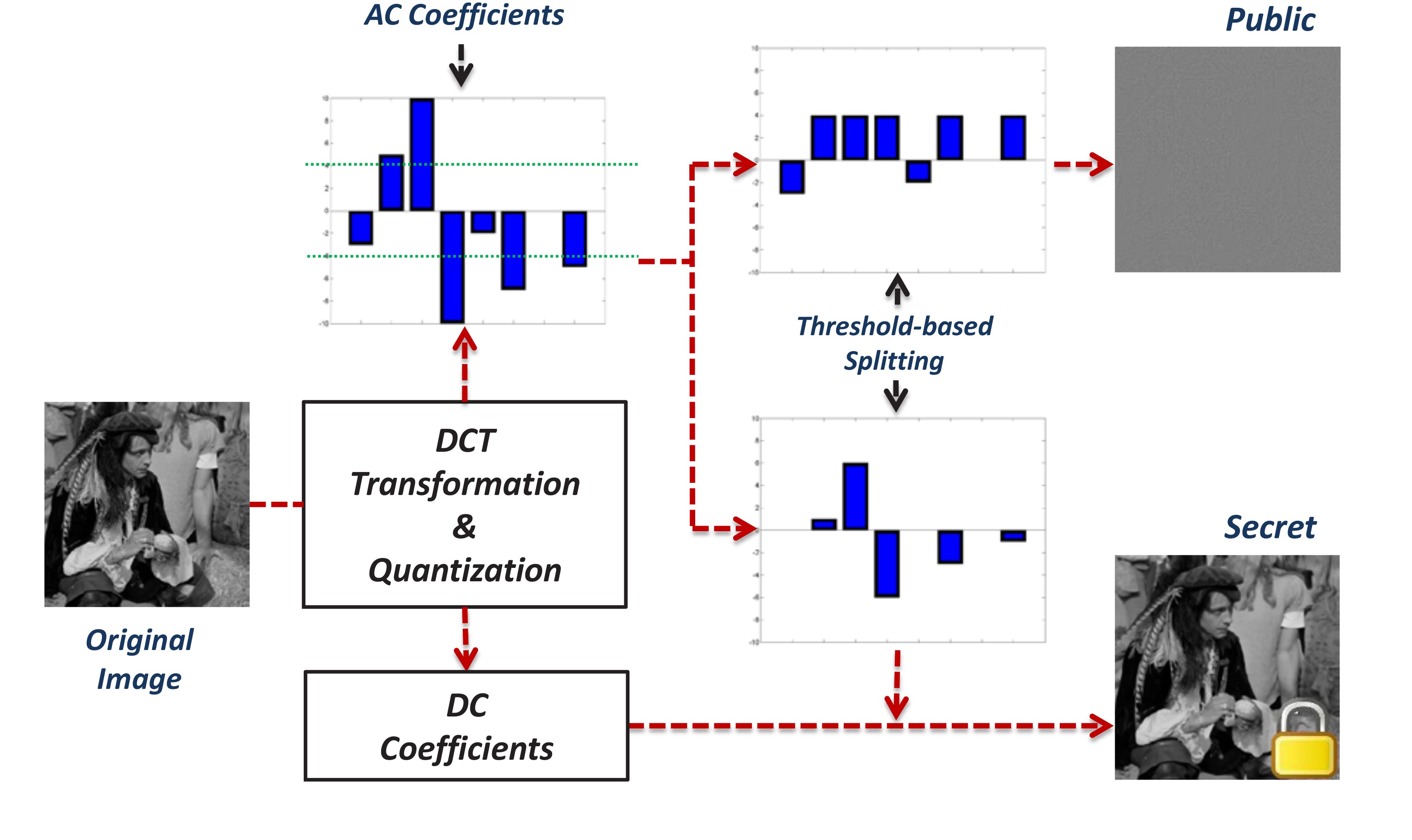}
  \caption{P3 works by removing the DC and large AC coefficients from the public
    version of image and placing them in a secret image. (Image from
    ~\cite{p3})}
 \label{fig:p3}
\end{figure}

To explain how P3 works, we first review the basics of JPEG image
formatting~\cite{jpeg}.  When converting an image to JPEG, the color values of
each block of $8\times8$ pixels are processed with a discrete cosine transform
(DCT), and the resulting DCT coefficients are saved as the image representation
(after some additional formatting).

P3 assumes that the DC coefficient (the $0^{th}$ one) and the larger AC
coefficients (the remaining $63$) carry the most information about the
image.  When applying P3 to an image, the user sets a \emph{threshold}
value.  The DC coefficients, as well as any AC coefficients whose values
are larger than the threshold, are encrypted and stored in the secret
image (see Figure~\ref{fig:p3}).  In the public image, these coefficients
are replaced by the threshold value.  All remaining AC coefficients are
stored in in the public image in plaintext.  To reconstruct the original
image, the coefficients in the secret image are decrypted and copied to
their places in the public image.

By varying the threshold value, the user can balance the size of the secret
image against the amount of data removed from the public image.  The authors of
P3 recommend using threshold values between 10 and 20.

P3 explicitly aims to protect images from ``automatic recognition
technologies''~\cite{p3}.  P3 uses signal processing tests, including
edge detection, face detection, face recognition, and SIFT features,
as privacy metrics.  Since these features in the public images do not
match those in the original images and standard techniques do not find or
recognize faces in the public images, the authors of P3 conclude that the
system protects privacy.  They do not evaluate the human recognition rate
of public images, but as examples in Table~\ref{tab:samples} illustrate,
public P3 images produced with the recommended threshold values do not
appear to be human-recognizable.

\section{Artificial neural networks}

Neural networks with many layers, also known as deep neural networks, have
become very effective in classifying high-dimensional data such as images
and speech signals~\cite{hannun2014deep, he2015delving, lecun2015deep}.
As opposed to other machine learning algorithms, which require explicit
feature specification and engineering, neural networks can, given a
classification task, \emph{automatically} extract complex features from
high-dimensional data and use these features to find the relationship
between the data and the model's output.  In this paper, we focus on
training neural networks in a supervised setting for classification.
The training process involves constructing a model that learns the
relationship between data and classes from a labeled dataset (where for
each data record we know its true class).

An artificial neural network is composed of multiple layers of nonlinear
functions so that more abstract features are computed as nonlinear
functions of lower-level features.  Each function is modeled as a neuron
in the network.  The input to the inner layer's neurons is the output
of the neurons in the previous layer.  The functions implemented in
the neurons are referred to as \emph{activation} units and are usually
pre-determined by the model designer.  The nonlinear functions are
constructed to maximize the prediction accuracy of the whole model on
a training dataset.  In Section~\ref{sec:networks}, we give concrete
examples of neural networks used for image recognition tasks.  

In our networks, we use two activation units: ReLU (rectified linear units) and
LeakyReLU.  The ReLU is an activation function $f(x) = \max(0, x)$, while the
LeakyReLU is the activation function
$$
f(x)=
\begin{cases}
    x,& \text{if } x > 1\\
    0.01 x,& \text{otherwise.}
\end{cases}
$$

Model training or learning is an optimization problem whose objective is
to (1) extract the most relevant features for a particular classification
task, and (2) find the best relationship between these features and the
output of the model (i.e., classification of inputs).

\begin{figure}[t]
  \captionsetup[subfigure]{justification=centering}
  \centering
  \includegraphics[width=\columnwidth]{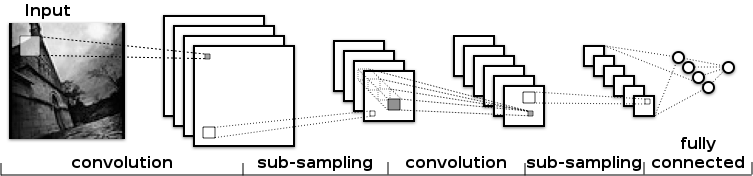}
  \caption{Schematic architecture of a convolutional neural
    network~\cite{convolutionimage}. The network is composed of convolutional
    layers followed by max-pooling sub-sampling layers.  The last layers are
    fully connected.}
 \label{fig:cnn}
\end{figure}

The architecture of the network, which determines how different
layers interact with each other, greatly affects the potential
accuracy of the model.  \emph{Convolutional neural networks}
(CNN) are a common deep neural-network architecture for object
recognition and image classification, known to yield high prediction
accuracy~\cite{krizhevsky2012imagenet, simonyan2014very}.  CNNs have also
been widely used for face recognition.  Zhou et al.~\cite{zhou2015naive}
and Parkhi et al.~\cite{parkhi2015deep} presented CNNs for the Labeled
Faces in the Wild database~\cite{huang2007labeled} supplemented with
large datasets of Web images.  Also, FaceNet~\cite{schroff2015facenet}
is an efficient CNN for facial recognition trained on triplets of two
matching and one non-matching facial thumbnails.

As opposed to fully connected neural networks, in which all neurons
in one layer are connected to all neurons in the previous layer, in
a convolutional neural network each neuron may process the output of
only a subset of other neurons.  This technique allows the model to
embed some known structure of the data (e.g., relationship between
neighboring pixels).  Figure~\ref{fig:cnn} illustrates the schematic
architecture of a convolutional neural network.  These networks are
composed of multiple convolutional and sub-sampling layers with several
fully connected layers at the end.  Each neuron in a convolutional layer
works on a small connected region of the previous layer (e.g., neighboring
pixels in the first layer) over their full depth (e.g., all color signals
in the first layer).  Usually, multiple convolutional neurons operate
on the same region in the previous layer, which results in extracting
multiple features for the same region.  In the sub-sampling layer, a
function of these features is computed and passed to the next layer.
Max-pooling, where the feature with the highest value is selected,
is typically used for this purpose.  After a few convolutional and
max-pooling layers, the final layers of the network are fully connected.

Given a network architecture, the training process involves learning
the importance weights of each input for each neuron.  These weights are
referred to as neural-network parameters.  The objective is to find the
parameter values that minimize the classification error of the model's
output.  Training is an iterative process.  In each iteration (known
as a training epoch), the weights are updated so as to reduce the model's
error on its training set.  This is done by stochastic gradient descent
algorithm~\cite{rumelhart1988learning} which computes the gradient of
the model's error over each parameter and updates the parameter towards
lower errors.  The magnitude of the update is controlled by the ``learning
rate'' hyper-parameter.

To avoid overfitting on the training set and to help the trained model
generalize to data beyond its training set, various regularization
techniques are used.  Dropout~\cite{srivastava2014dropout} is an effective
regularization technique for neural networks which works by randomly
dropping nodes (neurons) during the training with a pre-determined
probability.

\section{Threat model}
\label{sec:threat}

We assume that the adversary has a set of obfuscated images and his goal
is to uncover certain types of sensitive information hidden in these
images: namely, recognize objects, faces, and/or digits that appear in
the image.  Recognizing objects or faces from a known set is a standard
image recognition task, except that in this case it must be performed
on obfuscated images.  For example, the operator of a storage service
that stores obfuscated photos from an online social network may want to
determine which users appear in a given photo.

We also assume that the adversary has access to a set of plain,
unobfuscated images that can be used for training the adversary's neural
networks.  In the case of objects or handwritten digits, the adversary
needs many different images of objects and digits.  Such datasets are
publicly available and used as benchmarks for training image recognition
models.  In the case of face recognition, the adversary needs the set of
possible faces that may appear in a given photo.  This is a realistic
assumption for online social networks, where the faces of most users
are either public, or known to the network operator.

We assume that the adversary knows the exact algorithm used to
obfuscate the images but not the cryptographic keys (if any) used
during obfuscation.  In the case of P3, this means that the adversary
knows which threshold level was used but not the keys that encrypt the
significant JPEG coefficients.  In the case of mosaicing, the adversary
knows the size of the pixelation window.  In the case of blurring,
the adversary has \emph{black-box} access to the blurring algorithm
and does not have any information about this algorithm other than
what this algorithm produces on adversary-supplied videos and images.
This accurately models the case of YouTube blurring, which from the
viewpoint of a video creator has a simple on/off switch.

\section{Methodology}

\subsection{How the attack works}

The main idea of our attack is to train artificial neural networks
to perform image recognition tasks on obfuscated images.  We train a
separate neural-network model for each combination of an obfuscation
technique and a recognition task.

As explained in Section~\ref{sec:threat}, we assume that the adversary
has access to a set of plain, unobfuscated images that he can use
for training.  We generate the training set by applying the given
obfuscation technique to these images (for example, request YouTube's
Video Manager to blur the faces).  We then perform supervised learning on
the obfuscated images to create an obfuscated-image recognition model,
as described in Section~\ref{sec:training}.  Complete descriptions of
our neural-network architectures are in the appendices.  Finally, we
measure the accuracy of our models.

In all of our experiments, the training set and the test set are disjoint.
For example, the images used for training the mosaiced-face recognition
model are drawn from the same dataset of facial photos as the images
used for measuring the accuracy of this model, but the two subsets have
no images in common.

\subsection{Datasets}
\begin{table*}[t] 
  \centering
  \def \samp {43px}
  \begin{tabular}{p{40px}|m{\samp}|m{\samp}|m{\samp}|m{\samp}|m{\samp}|m{\samp}|m{\samp}|m{\samp}} 
    \hline 

    &
    &
    \multicolumn{4}{c|}{Mosaic} &
    \multicolumn{3}{c}{P3} \\
    
    Dataset &
    Original &
    \multicolumn{1}{c}{$2\times2$} & 
    \multicolumn{1}{c}{$4\times4$} & 
    \multicolumn{1}{c}{$8\times8$} & 
    \multicolumn{1}{c|}{$16\times16$} &
    \multicolumn{1}{c}{20} & 
    \multicolumn{1}{c}{10} & 
    \multicolumn{1}{c}{1} \\
    \hline

    MNIST & 
    \includegraphics[width=\samp]{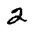} & 
    \includegraphics[width=\samp]{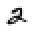} &
    \includegraphics[width=\samp]{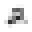} &
    \includegraphics[width=\samp]{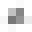} &
    \includegraphics[width=\samp]{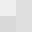} & 
    \includegraphics[width=\samp]{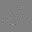} & 
    \includegraphics[width=\samp]{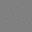} &
    \includegraphics[width=\samp]{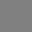} \\

    CIFAR-10 & 
    \includegraphics[width=\samp]{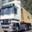} & 
    \includegraphics[width=\samp]{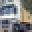} &
    \includegraphics[width=\samp]{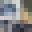} &
    \includegraphics[width=\samp]{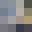} &
    \includegraphics[width=\samp]{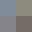} &
    \includegraphics[width=\samp]{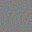} & 
    \includegraphics[width=\samp]{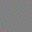} &
    \includegraphics[width=\samp]{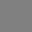} \\

    AT\&T & 
    \includegraphics[width=\samp]{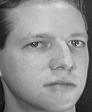} & 
    \includegraphics[width=\samp]{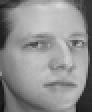} &
    \includegraphics[width=\samp]{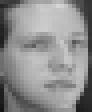} &
    \includegraphics[width=\samp]{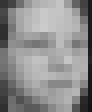} &
    \includegraphics[width=\samp]{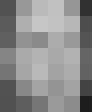} & 
    \includegraphics[width=\samp]{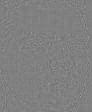} & 
    \includegraphics[width=\samp]{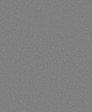} &
    \includegraphics[width=\samp]{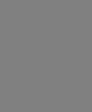} \\

    FaceScrub & 
    \includegraphics[width=\samp]{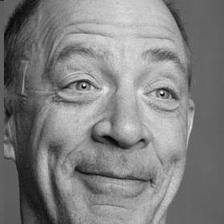} & 
    \includegraphics[width=\samp]{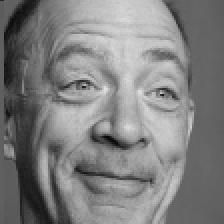} &
    \includegraphics[width=\samp]{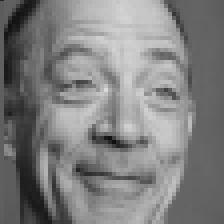} &
    \includegraphics[width=\samp]{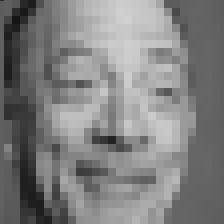} &
    \includegraphics[width=\samp]{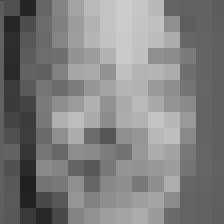} &
    \includegraphics[width=\samp]{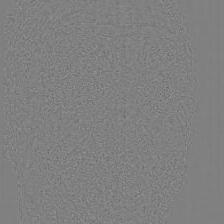} & 
    \includegraphics[width=\samp]{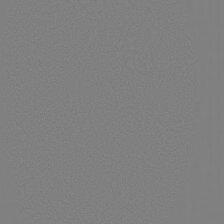} &
    \includegraphics[width=\samp]{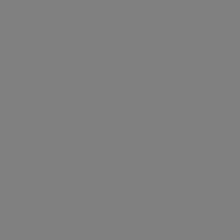} \\
    \hline
  \end{tabular}

  \caption{Examples images from each dataset. The leftmost image is the
    original image. The remaining columns are the image obfuscated with
    mosaicing with windows of $2\times2$, $4\times4$, $8\times8$, and
    $16\times16$ pixels and P3 with thresholds of 20, 10, and 1.}
  \label{tab:samples}
\end{table*}

We used four different, diverse datasets: the MNIST database of
handwritten digits, the CIFAR-10 image dataset, the AT\&T database of
faces, and the FaceScrub celebrity facial dataset.

\paragraphbe{MNIST.} 
The MNIST dataset~\cite{mnist} consists of $28\times28$ grayscale images
of handwritten digits collected from US Census Bureau employees and
high-school students. Each image consists of one handwritten digit
(i.e., an Arabic numeral between 0 and 9).  The dataset is divided
into a training set of 60,000 images and a test set of 10,000 images.
We expanded the MNIST images to 32x32 images by adding a 2-pixel white
border around the edge of each image.

\paragraphbe{CIFAR-10.} 
The CIFAR-10 dataset~\cite{cifar} consists of $32\times32$ color images.
Each image contains an object belonging to one of 10 classes.  Each class
is either a vehicle (e.g., plane, car, etc.) or an animal (e.g., dog,
cat, etc.). There are 50,000 images in the CIFAR-10 training set and
10,000 images in the test set.

\paragraphbe{AT\&T.} 
The AT\&T database of faces~\cite{att} contains 400 $92\times112$
grayscale images of 40 individuals.  Each individual has 10 images
in the dataset, taken under a variety of lighting conditions, with
different expressions and facial details (i.e., with or without glasses).
For our training set, we randomly selected 8 images of each person.
The remaining 2 were used in the test set.

\paragraphbe{FaceScrub.}  
The FaceScrub dataset~\cite{facescrub} is a large dataset originally
consisting of over 100,000 facial images of 530 celebrities.  The images
have been compiled from various online articles.  Due to copyright
concerns only the image URLs are distributed, not the images themselves.
Some of the images are no longer accessible and we were able to
download only 66,033 images. FaceScrub includes the bounding boxes
for the faces in each image and we used those to extract the faces. 10
images of each person was used in the test set, the remaining 60,733
were used for training.  Because some of the images are not in color,
we converted all images to grayscale and scaled them to $224\times224$.

\subsection{Obfuscation}
\label{sec:applyingobfuscation}

For mosaicing (pixelation), we used a simple Python script with
NumPy~\cite{oliphant2007python} to compute the average color of a block
of pixels and then change the entire block to that color.

To obfuscate images using the P3 technique, we modified the
9a version of the Independent JPEG Group's JPEG compression
software~\cite{independent2012group} to replace any JPEG coefficient whose
absolute value is bigger than the threshold with the threshold value.
The resulting image is the same as the public image that would have been
produced by P3.

For blurring faces in the AT\&T dataset, we used YouTube's facial
blurring~\cite{youtubeface}. For the training and test sets, we used
$\mathtt{ffmpeg}$ to create videos of the original faces from the dataset
and uploaded them to YouTube.  Each face was shown for 1 second and
followed by 1 second of white frames before the next face was shown.
Video resolution was $1280\times720$, with $92\times112$ faces were
centered in their frames.

After uploading the training and test videos, we used YouTube's Video
Manager to automatically blur all faces in these videos, then downloaded
the videos and used $\mathtt{ffmpeg}$ to extract each frame.  We did
not notice any faces that YouTube did not blur, but some edges of a few
images were not blurred.

Although the images in the videos were static, many of the faces in the
blurred videos shifted in appearance, that is, parts of an image would
become lighter or darker between frames.  We do not know if this is a
feature added by YouTube to make identification harder or an artifact
of their blurring technique.  To account for the different blurrings of
each image and to avoid any bleeding across images, we used the middle
5 frames of each image in the videos. This increased the size of our
training and testing sets to 1,600 and 400 images each.

\begin{figure}[tb]
  \captionsetup[subfigure]{justification=centering}
  \centering
  \begin{subfigure}{0.3\columnwidth}
    \centering
  \includegraphics[width=0.5\columnwidth]{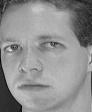}  
  \end{subfigure}%
  \begin{subfigure}{0.3\columnwidth}
    \centering
  \includegraphics[width=1\columnwidth]{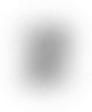}  
  \end{subfigure}%
  \begin{subfigure}{0.3\columnwidth}
    \centering
  \includegraphics[width=1\columnwidth]{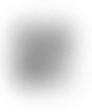}  
  \end{subfigure}
  \caption{An original AT\&T image and two blurred frames extracted from a
    blurred YouTube video. Although the unblurred frames were identical, the two
    blurred frames are different.}
 \label{fig:blurred}
\end{figure}

Because the blurring often extended outside of the original image borders,
we extracted the center $184\times224$ pixels from each frame and then
resized them to $92\times112$ pixels. Two examples of a blurred image
can be seen in Figure~\ref{fig:blurred}.

\subsection{Neural networks}
\label{sec:networks}

Our experiments were done with three different neural networks: a
digit recognition model for the MNIST dataset, an object recognition
model for CIFAR-10, and a face recognition model for AT\&T and
FaceScrub datasets.  All of our models are deep convolutional
neural networks \cite{jarrett2009best, krizhevsky2012imagenet} with
dropout~\cite{srivastava2014dropout} regularization.

Note that we use the same architecture for training classification models
on the original and obfuscated images.  We compare the accuracy of our
models on obfuscated images with the accuracy of similar models on the
original images.  If we had used more accurate neural-network models or
tailored neural networks specifically for recognizing obfuscated images,
we could have achieved even higher accuracy than reported in this paper.

\paragraphbe{MNIST.} 
We used a simple neural network for classifying images in the MNIST
dataset, based on Torch's template neural network for MNIST~\cite{torch}.
See Appendix~\ref{app:mnist} for the exact description of the network
architecture.

In this network, each convolutional layer with a leaky rectified linear
unit (LeakyReLU) is followed by a layer of pooling.  The network ends
with a simple fully connected layer and a softmax layer that normalizes
the output of the model into classification probabilities.  A dropout
layer with a probability of 0.5 is introduced between the linear layer
and the softmax layer.

\paragraphbe{CIFAR-10.} 
For CIFAR-10, we used Zagoruyko's CIFAR-10 neural
network~\cite{zagoruyko2015}.  This network was created to see how batch
normalization worked with dropout. Batch normalization speeds up the
training and improves neural networks by normalizing mini-batches between
layers inside the neural network~\cite{ioffe2015batch}.  The Zagoruyko
method creates a large convolutional neural network and adds batch
normalization after every convolutional and linear layer.  Zagoruyko's
network consists of 13 convolutional layers with batch normalization,
each with a rectified linear unit (ReLU).  See Appendix~\ref{app:cifar}
for the exact description of the network architecture.

\paragraphbe{AT\&T and FaceScrub.} 
The networks used for the AT\&T and FaceScrub datasets of facial images
are similar to the one used on the MNIST dataset, with an extra round
of convolution.   See Appendix~\ref{app:att} and~\ref{app:facescrub}
for the exact description of the network architecture.

\begin{table*}[tb]
  \centering
    \def \acc {0.059}
  \begin{tabular}{C{0.127}|C{0.063}|C{0.063}|L{\acc}|L{\acc}|L{\acc}|L{\acc}|L{\acc}|L{\acc}|L{\acc}}
    \hline			
    
    &
    &
    &
    \multicolumn{4}{c|}{Mosaic} &
    \multicolumn{3}{c}{P3} \\
    
    Dataset &
    Baseline &
    Original &
    \multicolumn{1}{c}{$2\times2$} & 
    \multicolumn{1}{c}{$4\times4$} & 
    \multicolumn{1}{c}{$8\times8$} & 
    \multicolumn{1}{c|}{$16\times16$} &
    \multicolumn{1}{c}{20} & 
    \multicolumn{1}{c}{10} & 
    \multicolumn{1}{c}{1} \\
    \hline

    MNIST Top 1 &
    10.00 &
    98.71 &
    98.49 & 96.17 & 83.42 & 52.13 &
    79.93 & 74.19 & 58.54 \\

    \hline

    MNIST Top 5 &
    50.00 &
    100 &
    100 & 99.95 & 99.36 & 93.90 &
    98.91 & 97.95 & 94.82 \\
    \hline
    \hline

    CIFAR Top 1 &
    10.00 &
    89.57 &
    81.76 & 70.21 & 53.95 & 31.81 &
    74.56 & 65.98 & 33.21 \\
    \hline

    CIFAR Top 5 &
    50.00 &
    99.46 & 
    98.85 & 97.10 & 92.26 & 81.76 &
    96.98 & 94.99 & 80.72 \\
    \hline
    \hline

    AT\&T Top 1 &
    2.50 &
    95.00 &
    95.00 & 96.25 & 95.00 & 96.25 &
    97.50 & 93.75 & 83.75 \\
    \hline

    AT\&T Top 5 &
    12.50 &
    100 &
    100 & 100 & 98.75 & 98.75 &
    100 & 100 & 95.00 \\
    \hline
    \hline

    FaceScrub Top 1 &
    0.19 &
    75.49 &
    71.53 & 69.91 & 65.25 & 57.56 &
    40.02 & 31.21 & 17.42 \\
    \hline

    FaceScrub Top 5 &
    0.94 &
    86.06 & 
    83.74 & 82.08 & 79.13 & 72.23 & 
    58.38 & 51.28 & 34.79 \\
    \hline
  \end{tabular}
  \caption{Accuracy of neural networks classifying the original datasets as well
    as those obfuscated with mosaicing with windows of $2\times2$, $4\times4$,
    $8\times8$, and $16\times16$ pixels and P3 thresholds of 20, 10, and 1. The
    baseline accuracy corresponds to random guessing.}
  \label{tab:results}
\end{table*}

\subsection{Training}
\label{sec:training}

For each of our experiments, we obfuscated the entire dataset and then
split it into a training set and a test set.  In the MNIST and CIFAR-10
datasets, images are already designated as training or test images.
For AT\&T and FaceScrub, we randomly allocated images to each set.

For training the MNIST model, we used the learning rate of 0.01
with the learning rate decay of $10^{-7}$, momentum of 0.9, and
weight decay of $5\times10^{-4}$.  The learning rate and momentum
control the magnitude of updates to the neural-network parameters
during the training~\cite{rumelhart1988learning, bishop2006pattern}.
For training the CIFAR-10 model, we initialized the learning rate to $1$
and decreased it by a factor of $2$ every $25$ epochs.  Weight decay was
$5\times10^{-4}$, momentum was 0.9, and learning rate decay was $10^{-7}$.
For the AT\&T and FaceScrub models, we used the same learning rate and
momentum as in the MNIST training.

We ran all of our experiments for 100-200 training epochs.  For each
epoch, we trained our neural networks on the obfuscated training set
and then measured the accuracy of the network on the obfuscated test set.

Our neural networks were programmed in Torch. The MNIST and AT\&T networks
were distributed across multiple Linux machines in an HTCondor cluster.
The larger CIFAR-10 and FaceScrub networks made use of the Torch CUDA
backend and were trained on Amazon AWS g2.8xlarge machines with GRID
K520 Nvidia cards running Ubuntu 14.04.

\section{Results}

From each of the original datasets, we created seven obfuscated datasets
for a total of eight datasets (see Table~\ref{tab:samples}). For each
neural networks defined in~\ref{sec:networks}, we created eight models:
one for classifying images in the original dataset and one each for
classifying the obfuscated versions of that dataset. In addition to these
eight sets, we created a ninth set from the AT\&T dataset.  This set used
facial images that were blurred by YouTube.  While the same network was
used for all versions of a dataset, the networks were trained and tested
on only one version at a time (i.e., there was no mixing between the
images obfuscated with different techniques or with the original images).

Three of the obfuscated datasets were created by running P3 on the
original images and saving the P3 public images. We used P3 with
thresholds of 1, 10, and 20. 10 and 20 are the thresholds recommended
by the designers of P3 as striking a good balance between privacy and
utility~\cite{p3}.  The threshold of 1 is the most privacy-protective
setting that P3 allows.

The remaining four obfuscated datasets were created by mosaicing the
original dataset with different windows. Mosaic windows of $2\times2$,
$4\times4$, $8\times8$, $16\times16$ were used.  When analyzing the
accuracy of our network with different mosaic window sizes, image
resolution should be taken into account.  For datasets with $32\times32$
resolution (i.e., MNIST and CIFAR-10), $16\times16$ windows reduce the
practical size of each image to only $2\times2$ pixels.  On the FaceScrub
dataset, however, we could maintain the resolution of $14\times14$
pixels after the image was mosaiced with the same $16\times16$ window.

We computed the accuracy of the network in classifying the test set after
every round of training.  We recorded the accuracy of the top guess as
well as the top 5 guesses.  Results are shown in Table~\ref{tab:results}.

\begin{figure}[t]
  \captionsetup[subfigure]{justification=centering}
  \centering
  \begin{subfigure}{0.99\columnwidth}
    \centering
    \includegraphics[width=\columnwidth]{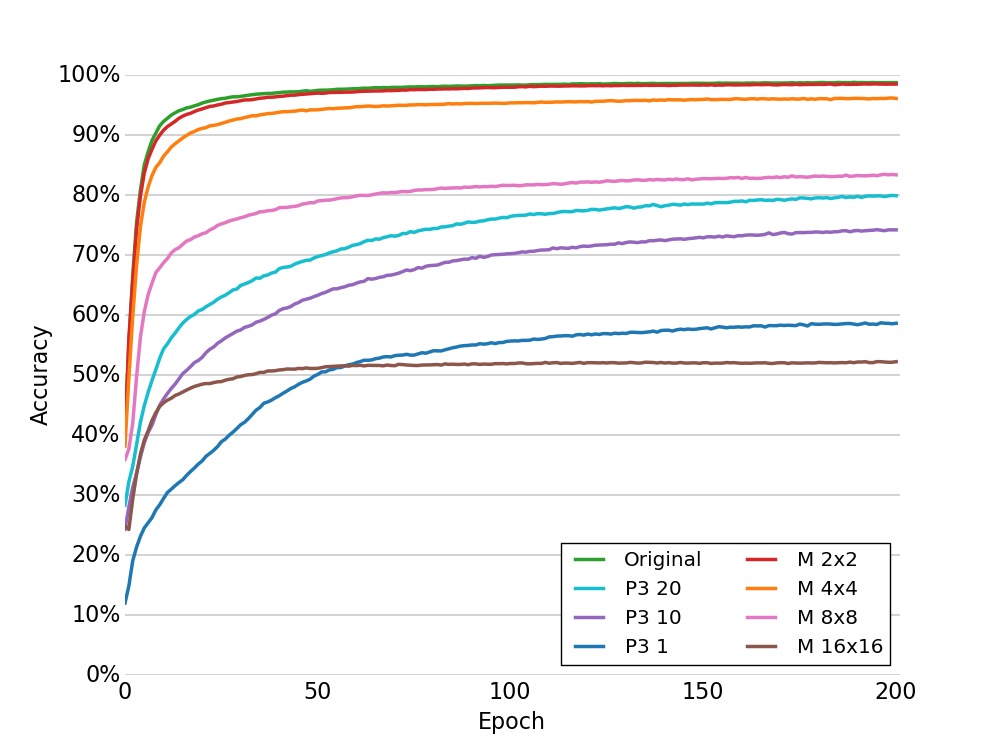}
    \caption{Top guess.}
  \end{subfigure}\\
  \begin{subfigure}{0.99\columnwidth}
    \centering
    \includegraphics[width=\columnwidth]{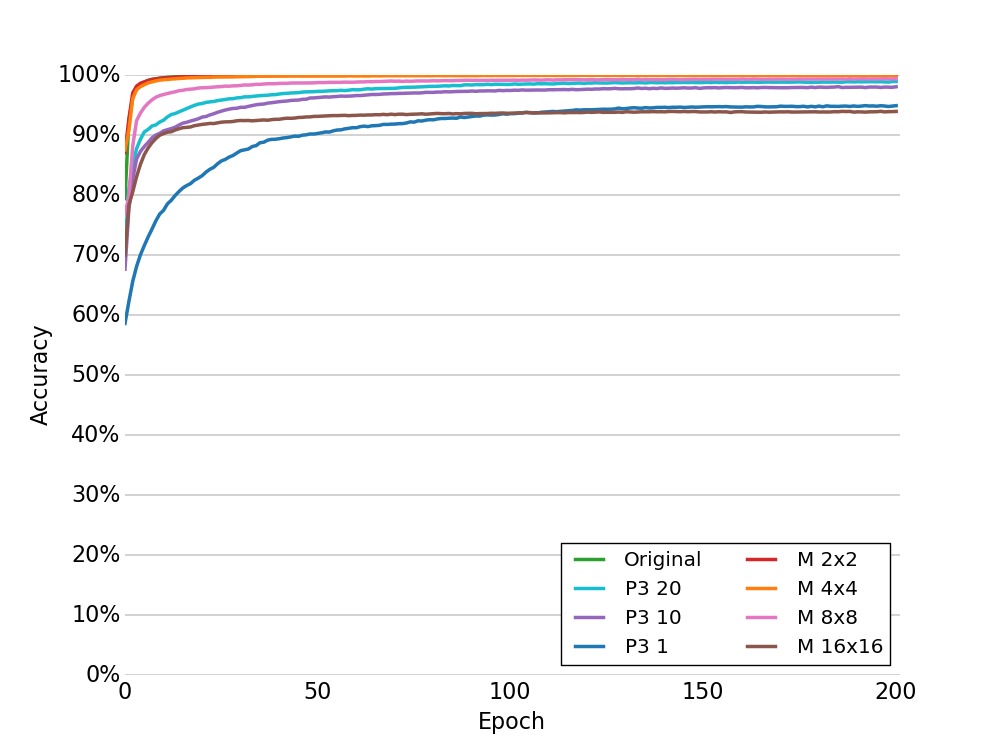}
    \caption{Top 5 guesses.}
  \end{subfigure}
  \caption{Test accuracy of the neural networks trained on the
    \textbf{MNIST} handwritten digits. The networks were trained
    and tested on digits obfuscated with different techniques: P3
    with thresholds of 1, 10, and 20, and mosaicing with $2\times2$,
    $4\times4$, $8\times8$, and $16\times16$ windows.}
 \label{fig:mnist-results}
\end{figure}

\subsection{MNIST}

\begin{figure}[t]
  \captionsetup[subfigure]{justification=centering}               
  \centering
  \begin{subfigure}{0.99\columnwidth}
    \centering
    \includegraphics[width=\columnwidth]{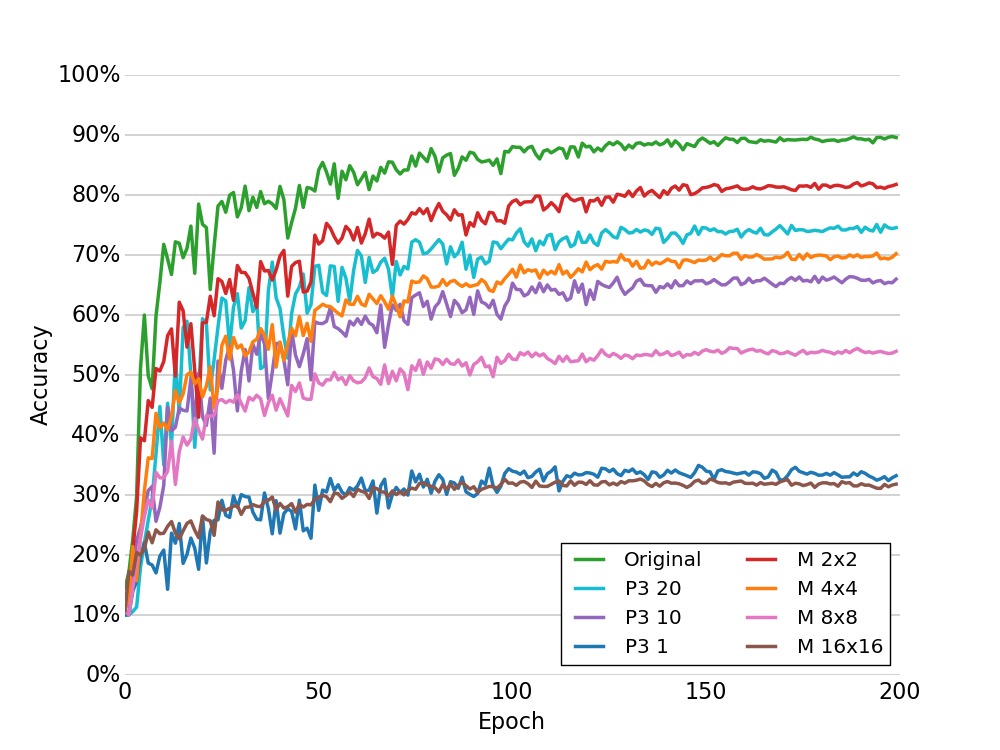}
    \caption{Top guess.}
  \end{subfigure}\\
  \begin{subfigure}{0.99\columnwidth}
    \centering
    \includegraphics[width=\columnwidth]{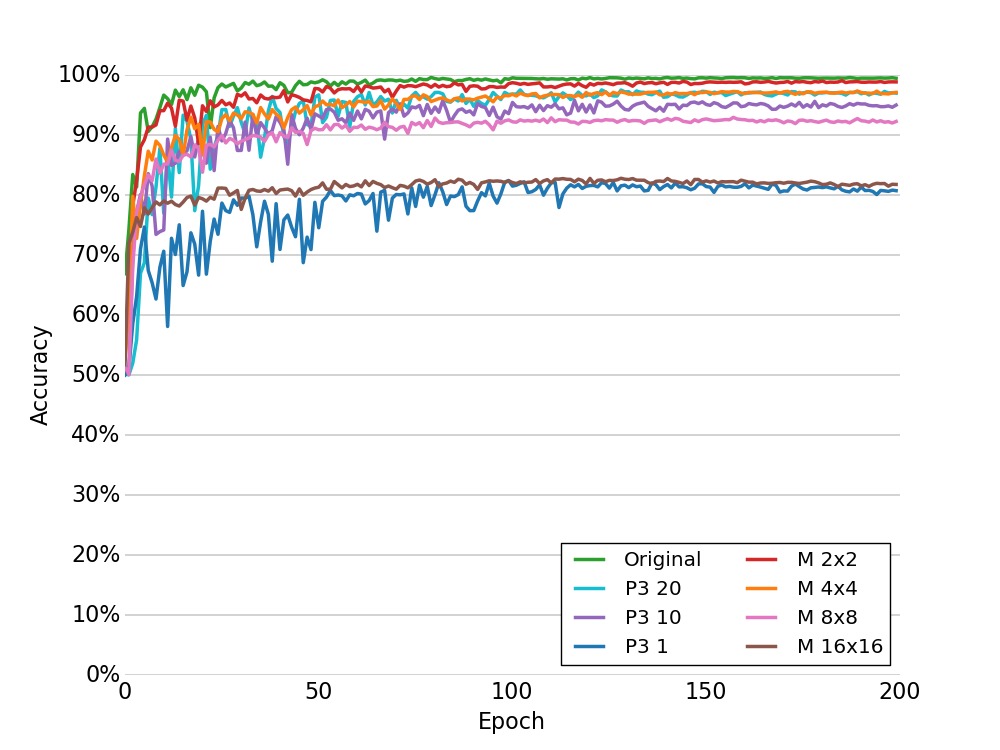}
    \caption{Top 5 guesses.}
  \end{subfigure}
 \caption{Test accuracy of the neural networks trained on the \textbf{CIFAR-10}
   images. The networks were trained and tested on colored images obfuscated
    with different techniques: P3
    with thresholds of 1, 10, and 20, and mosaicing with $2\times2$,
    $4\times4$, $8\times8$, and $16\times16$ windows.}
  \label{fig:cifar-results}
\end{figure}

The results for the MNIST neural network are shown in
Figure~\ref{fig:mnist-results}. The accuracy of the neural network
on the original images increases quickly and exceeds 90\% within 10
epochs. This is not surprising since MNIST is one of the older machine
learning datasets and is used pervasively to test models.  Top models
achieve 98\%-99\% accuracy~\cite{topclasses} and neural networks that
can get over 90\% accuracy are so simple that they are often used in
deep-learning and neural-network tutorials.

The models for mosaiced images with smaller windows (i.e., $2\times2$
and $4\times4$) also quickly exceeded 90\% accuracy. Although the MNIST
images are relatively small, just $32\times32$ pixels, these small
windows have little effect on obscuring the digits. The $2\times2$
mosaiced images are human-recognizable (see Table~\ref{tab:samples})
and the $4\times4$ mosaiced images still show the general shape and
pixel intensity to a large enough resolution that a neural network can
achieve accuracy of over 96\%.

The models for the $8\times8$ and $16\times16$ mosaiced images reached
accuracy of over 80\% and 50\%, respectively.  While these are not as
impressive as the other results, it's important to note that mosaicing
with these windows reduced the MNIST images to just $4\times4$ and
$2\times2$ unique pixels.  Even the accuracy of 50\% is significantly
larger than the 10\% chance of random guessing the correct digit.

The accuracy of recognizing public P3 images falls between the $8\times8$
and $16\times16$ mosaicing.  The accuracy of the threshold-20 model is
just below 80\%.  Looking at the threshold-20 image, the general shape of
the digit can be seen.  It is not surprising that the accuracy is close
to to the $8\times8$ mosaicing because P3 follows the JPEG specifications
and obfuscates each $8\times8$ block of pixels separately~\cite{jpeg}.

\subsection{CIFAR-10}

The CIFAR-10 model trained on the original images achieved just under
90\% accuracy. This is not as high as the MNIST results, but the CIFAR-10
images are much more complex and cluttered than the simple digits from
the MNIST dataset.  The CIFAR-10 mosaiced results are also not as strong
as the MNIST results. While it would seem that the larger amounts of
color information would make classification of the original and mosaiced
information easier, it also increases the dimensionality and complexity of
both the data and the neural network.  When using $16\times16$ windows,
the obfuscated CIFAR-10 images are reduced to just four colors.  It is
impressive that even in this challenging scenario, the accuracy of our
neural network is 31\%.

The P3 models on threshold-20 and threshold-10 images achieved accuracies
of 74\% and 66\%, respectively.  The accuracy on threshold-1 images,
however, dropped to only 32\%.

\begin{figure}[tb]
  \captionsetup[subfigure]{justification=centering}       
  \centering
  \begin{subfigure}{0.99\columnwidth}
    \centering
    \includegraphics[width=\columnwidth]{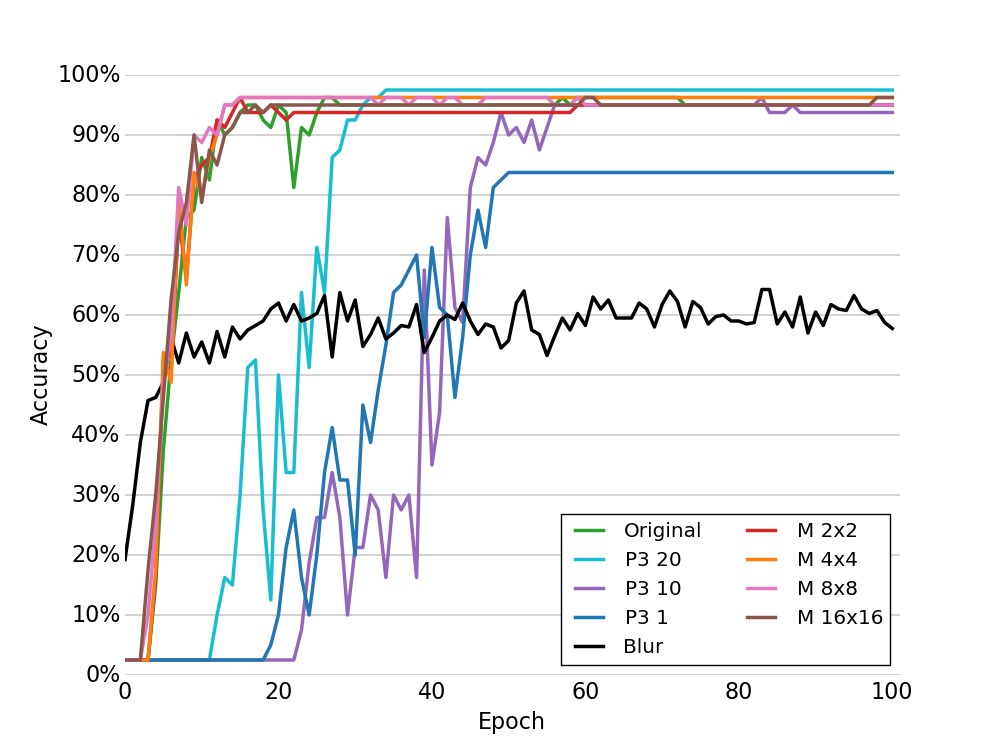}
    \caption{Top guess.}
  \end{subfigure}\\
  \begin{subfigure}{0.99\columnwidth}
    \centering
    \includegraphics[width=\columnwidth]{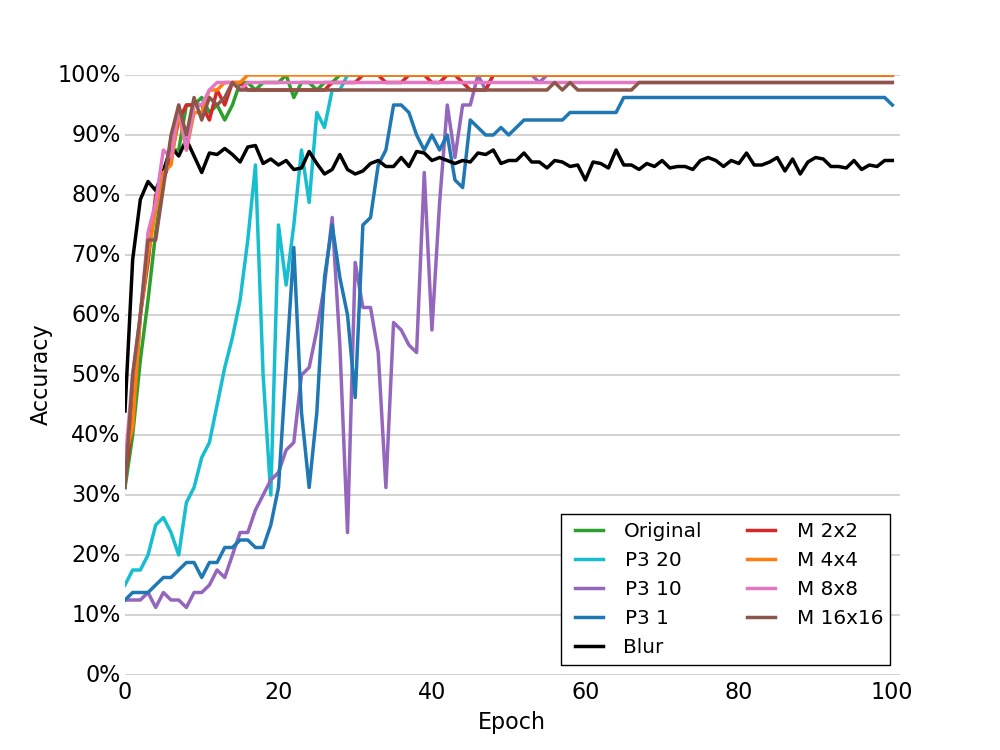}
    \caption{Top 5 guesses.}
  \end{subfigure}
  \caption{Test accuracy at each epoch of neural-network training on the
    \textbf{AT\&T} dataset of faces. The networks were trained and tested on
    black-and-white faces obfuscated with different techniques: P3 with
    thresholds of 1, 10, and 20, mosaicing with $2\times2$, $4\times4$,
    $8\times8$, and $16\times16$ windows, and automatic YouTube blurring.}
  \label{fig:att-results}
\end{figure}

\begin{figure}[tb]
  \captionsetup[subfigure]{justification=centering}       
  \centering
  \begin{subfigure}{0.99\columnwidth}
    \centering
    \includegraphics[width=\columnwidth]{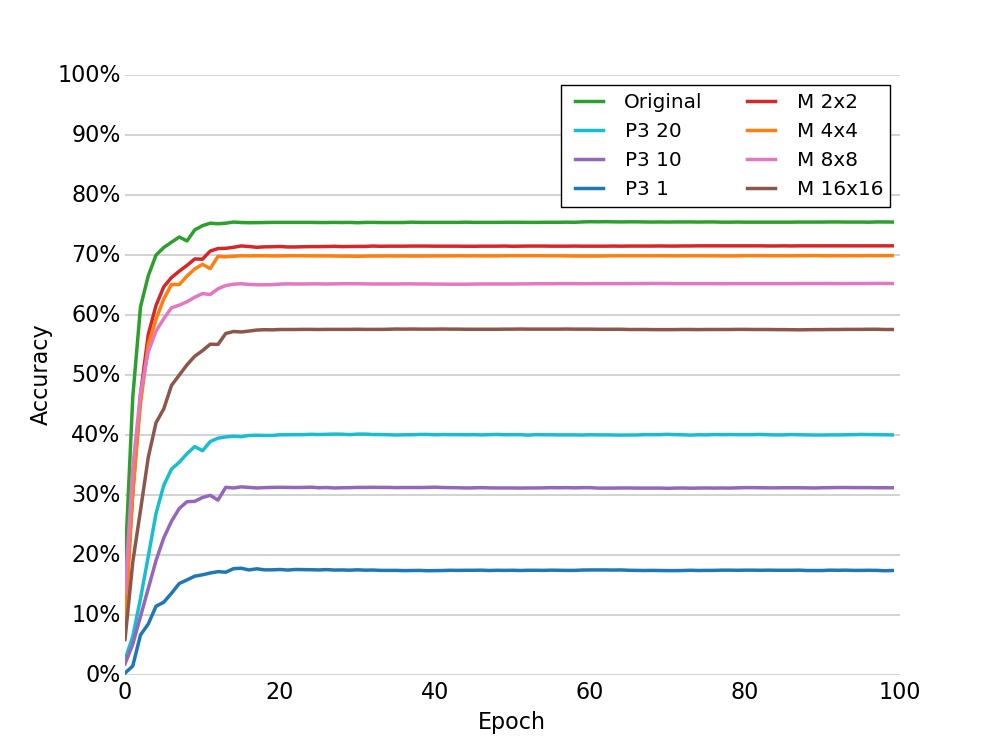}
    \caption{Top guess.}
  \end{subfigure}\\
  \begin{subfigure}{0.99\columnwidth}
    \centering
    \includegraphics[width=\columnwidth]{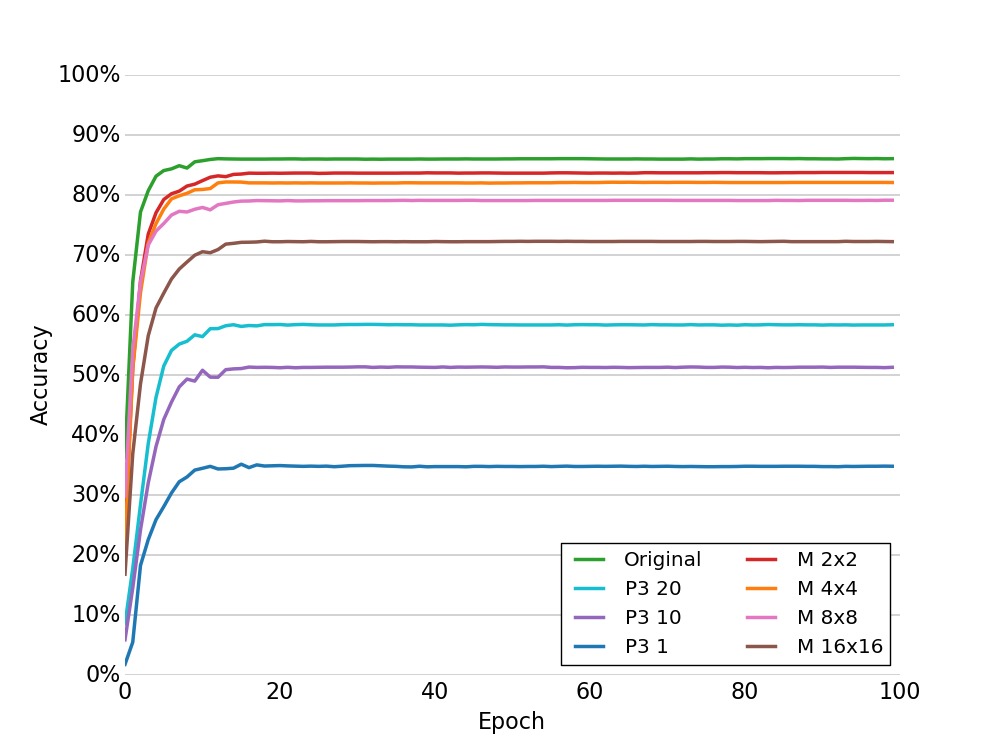}
    \caption{Top 5 guesses.}
  \end{subfigure}
  \caption{Test accuracy at each epoch of neural-network training on the
    \textbf{FaceScrub} dataset.  The networks were trained and tested on
    black-and-white celebrity faces obfuscated with different techniques:
    P3 with thresholds of 1, 10, and 20, mosaicing with $2\times2$,
    $4\times4$, $8\times8$, and $16\times16$ windows, and automatic
    YouTube blurring.}
  \label{fig:facescrub-results}
\end{figure}

\subsection{AT\&T}

\begin{table}[tb]
  \centering
  \begin{tabular}{c|c|c|c}
    \hline
    
    Dataset &
    Baseline &
    Original &
    Blurred \\
    \hline

    AT\&T Top 1 &
    2.50 &
    95.00 &
    57.75 \\

    AT\&T Top 5 &
    12.50 &
    100 &
    85.75 \\

    \hline
  \end{tabular}
  \caption{The accuracy of classifying the \textbf{AT\&T} faces blurred with
    YouTube.}
  \label{tab:youtube-acc}
\end{table}

The results for the models trained on the AT\&T dataset of faces are
shown in Figure~\ref{fig:att-results}. The models for the original and
mosaiced images all achieved over 95\% accuracy. For the original images
and the smaller mosaicing windows, this is not surprising. The images in
the AT\&T dataset are $92\times112$ pixels, much larger than the MNIST
and CIFAR-10's resolution of $32\times32$. Even the $8\times8$ mosaiced
faces are probably recognizable by a human (see Figure~\ref{tab:samples}).

A human might be able to recognize $16\times16$ mosaiced images as
faces, but we hypothesize that individual identification would become
challenging at that level of pixelization. However, these mosaiced images
have $6\times7$ resolution and there is still enough information for the
neural networks to be able to accurately recognize people in the images.

The models trained on P3 images did not all reach as high an accuracy as the
models working against mosaicing, but their accuracy was above 80\%, still much
higher than the 2.5\% accuracy of random guessing.  The results for the
threshold-20 images are the best, producing correct identification 97\% of the
time.  Looking at the threshold-20 images, rough outlines of faces can be seen,
although many of the helpful details, such as differences in color, have been
removed by the P3 algorithm. The accuracy of our model was lowest on the
threshold-1 images. However, it was still accurate over 83\% of the time, which
is a major improvement vs.\ the 2.5\% success rate of random guessing.

The AT\&T dataset was the only set that we obfuscated with YouTube's facial
blurring. Anecdotally, the authors of this paper were at a complete loss when
trying to identify the blurred faces by sight.  Our simple neural network,
however, was able to recognize individuals with 57.75\% accuracy
(Table~\ref{tab:youtube-acc}).

\subsection{FaceScrub}

The AT\&T dataset is relatively small, with only 40 individuals and 10
images per individual.  While it is helpful to illustrate the power of
neural networks, a larger dataset like FaceScrub, with 530 individuals,
is more indicative of achievable accuracy.  The full results of the
FaceScrub models are shown in Figure~\ref{fig:facescrub-results}.

The accuracy of our neural network on the original FaceScrub dataset
is 75\%.  This is impressive for such a simple network on a large
dataset. Once again, a more sophisticated network architecture could
likely achieve much better results, but our experiments show that even
a simple network can defeat the image obfuscation techniques.

The models for recognizing mosaiced faces exhibit the same pattern as the models
on the MNIST and CIFAR-10 datasets, with the smaller mosaicing window resulting
in the almost the same accuracy as the original images.  It is not surprising
that the accuracy of recognizing mosaiced faces did not drop below 50\%. The
FaceScrub images have relatively large resolution, $224\times224$, thus a
$16\times16$ window only reduces the resolution of the image to $14\times14$
pixels.

The results for FaceScrub protected using P3 show some of the worst
accuracies in all our experiments.  Nevertheless, even the threshold-1
accuracy of 17\% is still almost two orders of magnitude larger the
accuracy of random guessing (0.19\%).

\section{Related work}
\label{sec:related}

We survey related work in two areas: image obfuscation and applications
of neural networks in security and privacy.

\paragraphbe{Image obfuscation.}
Many existing systems and techniques use blurring and mosaicing to
protect users' privacy.  Face/Off~\cite{ilia2015face} uses facial
blurring to prevent identification in ``restricted'' Facebook
images.  YouTube supports blurring of faces~\cite{youtubeface} and
objects~\cite{youtubeblur} in videos as well.  Google Street View 
blurs license plates and faces~\cite{frome2009large}.

There is a large body of literature on why simple blurring
techniques are insufficient for protecting privacy.  To evaluate the
effectiveness of blurring and pixelation for hiding identities, Lander
et al.~\cite{lander2001evaluating} asked participants to identify famous
people in obfuscated movie clips and static images.  Their result show
that famous people can still be recognized in some obfuscated images.
They also reported higher identification accuracy for movie clips compared
to static images.  Neustaedter et al.~\cite{neustaedter2006blur} analyzed
how obfuscation can balance privacy with awareness in always-on remote
video situations, where users would like to recognize the remote person
without being able to see the privacy-sensitive parts of the video.
Their results show that blurring and pixelation do not achieve both
sufficient privacy and satisfactory awareness in remote work settings.

Newton et al.~\cite{newton2005preserving} examined many de-identification
techniques for obfuscated face images and achieved an extremely high
(99\%) recognition rate.  However, they only considered obfuscating a
small rectangle on the top part of the face, including the eyes and top
of the nose.  Gross et al.~\cite{gross2006model} designed an algorithmic
attack to identify people from their pixelated and blurred face images.
Their attack is based on the similarity of the obfuscated image and
the original images.  They showed that small mosaic boxes (e.g., 2-4
pixels) and simple blurring would not prevent identification attacks.
The authors suggested a new de-identification technique based on Newton et
al.~\cite{newton2005preserving}.  Cavedon et al.~\cite{cavedon2011getting}
exploited the changes in pixel boxes for obfuscating video frames to
reconstruct pixelated videos using image processing methods so that
humans can identify objects in the reconstructed images.  Wilber et
al.~\cite{wilber2016can} used Facebook's face tagging system as a
black-box tool to determine if faces obfuscated with various techniques,
including blurring, are still recognizable by Facebook.

Venkatraman~\cite{venkatraman2014} presented a brute-force attack to
reconstruct pixelated check numbers and concluded that obfuscating
sensitive information using blurring provides poor protection, although
it might not be easy to reconstruct faces from blurred images.  Hill et
al.~\cite{hill2016effectiveness} showed that obfuscated text can be
reconstructed with a large accuracy using language models.  They used
hidden Markov models (HMM) to achieve better speed and accuracy vs.
Venkatraman's method.  They were also able to accurately reconstruct
texts when the size of the mosaic box exceeds the size of the letters.
Hill's technique relies heavily on knowing the text font and size,
as well as the size of the mosaic box.

Gopalan et al.~\cite{gopalan2012blur} presented a method to recognize
faces obfuscated with non-uniform blurring by examining the space spanned
by the blurred images. Punnappurath et al.~\cite{punnappurath2015face}
extended this work by applying possible blurring effects to images in
the target gallery and finding the minimal distance between the gallery
images and the blurred image.

\paragraphbe{Neural networks.}
Neural networks have been successfully used extract information from
(unobfuscated) images.  For example, Golomb et al.~\cite{golomb1990sexnet}
used neural networks to identify the sex of human faces.  

Beyond their applications in image recognition, (deep) neural
networks have been used in many privacy and security contexts.
For example, Cannady et al.~\cite{cannady1998artificial} and Ryan
et al.~\cite{ryan1998intrusion} used neural networks in intrusion
detection systems (IDS).  Neural networks are particularly useful
for this purpose because the IDS designer does not need to engineer
relevant network-flow features and can rely on the network to discover
these features automatically.  Deep neural networks have been used for
malware classification in general~\cite{dahl2013large,pascanu2015malware}
and for specifically detecting Android malware~\cite{yuan2014droid}.

Deep convolutional neural networks can be used to detect objects
in images.  This enables detecting sensitive objects in a video
stream or static images.  Korayem et al.~\cite{korayem2016enhancing}
proposed a technique to detect computer screens in images.  The goal
is to alert the user or to hide the screen to protect privacy of
users who may have sensitive information on their screens.  Tran et
al.~\cite{tran2016privacy} detect privacy-sensitive photos using deep
neural networks.

Sivakorn et al.~\cite{sivakorn2016robot} used deep convolutional
neural networks to break Google's image CAPTCHAs (which ask users to
identify images that have a particular object in them).  Melicher et
al.~\cite{melicher2016fast} used deep neural networks to analyze user
passwords and construct a password prediction model.

Concurrently with our work, Oh et al.\ released a
preprint~\cite{oh2016faceless} where they use a neural network
to recognize untagged individuals in social-media images.  Our work
differs in several ways: 1) Oh et al.\ only examine the obfuscation of
faces in larger images while we work with entirely obfuscated images
(including backgrounds); 2) Oh et al.\ take advantage of unobfuscated
body cues and contextual information in the images to correlate multiple
images, whereas we do not make use of any external information beyond
the obfuscated image itself; 3) we focus on a broader class of image
recognition problems and defeat more types of obfuscation (including,
in the case of P3, partially encrypted images that are not recognizable
by humans, and real-world protections deployed by popular systems such as
YouTube); and 4) Oh et al.\ only evaluate a single dataset (the People
in Photo Albums~\cite{zhang2015beyond}), whereas we evaluate our attack
against diverse datasets, including MNIST, CIFAR-10, AT\&T Database of
Faces, and FaceScrub.  Considering the most comparable results (their
unary model of blurred faces across events vs.\ our blurred faces from
the AT\&T datasets), our model achieved 18\% higher accuracy than theirs.

\section{Conclusions}

The experiments in this paper demonstrate a fundamental problem faced
by ad hoc image obfuscation techniques.  These techniques partially
remove sensitive information so as to render certain parts of the image
(such as faces and objects) unrecognizable by humans, but retain the
image's basic structure and appearance and allow conventional storage,
compression, and processing.  Unfortunately, we show that obfuscated
images contain enough information correlated with the obfuscated content
to enable accurate reconstruction of the latter.

Modern image recognition methods based on deep learning are especially
powerful in this setting because the adversary does not need to specify
the relevant features of obfuscated images in advance or even understand
how exactly the remaining information is correlated with the hidden
information.  We demonstrate that deep learning can be used to accurately
recognize faces, objects, and handwritten digits even if the image has
been obfuscated by mosaicing, blurring, or encrypting the significant
JPEG coefficients.

Instead of informal arguments based on human users' inability to
recognize faces and objects in the obfuscated image, the designers of
privacy protection technologies for visual data should measure how
much information can be reconstructed or inferred from the obfuscated
images using state-of-the-art image recognition algorithms.  As the
power of machine learning grows, this tradeoff will shift in favor of
the adversaries.  At the other end of the spectrum, full encryption
blocks all forms of image recognition, at the cost of destroying all
utility of the images.  

How to design privacy protection technologies that can, for example,
protect faces in photos and videos while preserving the news value of
these images is an important topic for future research.

\bibliographystyle{abbrv}
{\small \bibliography{bib}}

\begin{thebibliography}{10}

\bibitem{warondrugs}
D.~Aitkenhead.
\newblock {`I've done really bad things': The undercover cop who abandoned the
  war on drugs}.
\newblock {\em The Guardian}, 2016.

\bibitem{att}
{AT\&T Laboratories Cambridge}.
\newblock The database of faces, 1994.

\bibitem{topclasses}
R.~Benenson.
\newblock Who is the best at {X}?
\newblock \url{http://rodrigob.github.io/are_we_there_yet/build/}, 2016.

\bibitem{bishop2006pattern}
C.~M. Bishop.
\newblock Pattern recognition.
\newblock {\em Machine Learning}, 128, 2006.

\bibitem{cannady1998artificial}
J.~Cannady.
\newblock Artificial neural networks for misuse detection.
\newblock In {\em National information systems security conference}, 1998.

\bibitem{cavedon2011getting}
L.~Cavedon, L.~Foschini, and G.~Vigna.
\newblock Getting the face behind the squares: Reconstructing pixelized video
  streams.
\newblock In {\em WOOT}, 2011.

\bibitem{torch}
R.~Collobert, K.~Kavukcuoglu, and C.~Farabet.
\newblock Torch7: A matlab-like environment for machine learning.
\newblock In {\em BigLearn, NIPS Workshop}, 2011.

\bibitem{dahl2013large}
G.~E. Dahl, J.~W. Stokes, L.~Deng, and D.~Yu.
\newblock Large-scale malware classification using random projections and
  neural networks.
\newblock In {\em 2013 IEEE International Conference on Acoustics, Speech and
  Signal Processing}, 2013.

\bibitem{convolutionimage}
Deeplearning.net.
\newblock Convolutional neural networks ({LeNet}).
\newblock \url{http://deeplearning.net/tutorial/lenet.html}, 2016.

\bibitem{frome2009large}
A.~Frome, G.~Cheung, A.~Abdulkader, M.~Zennaro, B.~Wu, A.~Bissacco, H.~Adam,
  H.~Neven, and L.~Vincent.
\newblock Large-scale privacy protection in {Google Street View}.
\newblock In {\em 2009 IEEE 12th international conference on computer vision},
  2009.

\bibitem{golomb1990sexnet}
B.~A. Golomb, D.~T. Lawrence, and T.~J. Sejnowski.
\newblock {SEXNET}: A neural network identifies sex from human faces.
\newblock In {\em NIPS}, 1990.

\bibitem{gopalan2012blur}
R.~Gopalan, S.~Taheri, P.~Turaga, and R.~Chellappa.
\newblock A blur-robust descriptor with applications to face recognition.
\newblock {\em IEEE transactions on pattern analysis and machine intelligence},
  2012.

\bibitem{gross2006model}
R.~Gross, L.~Sweeney, F.~De~la Torre, and S.~Baker.
\newblock Model-based face de-identification.
\newblock In {\em CVPRW}, 2006.

\bibitem{hannun2014deep}
A.~Hannun, C.~Case, J.~Casper, B.~Catanzaro, G.~Diamos, E.~Elsen, R.~Prenger,
  S.~Satheesh, S.~Sengupta, A.~Coates, et~al.
\newblock Deep speech: Scaling up end-to-end speech recognition.
\newblock {\em arXiv:1412.5567}, 2014.

\bibitem{he2015delving}
K.~He, X.~Zhang, S.~Ren, and J.~Sun.
\newblock Delving deep into rectifiers: Surpassing human-level performance on
  {ImageNet} classification.
\newblock In {\em Proceedings of the IEEE International Conference on Computer
  Vision}, 2015.

\bibitem{hill2016effectiveness}
S.~Hill, Z.~Zhou, L.~Saul, and H.~Shacham.
\newblock On the (in) effectiveness of mosaicing and blurring as tools for
  document redaction.
\newblock {\em PETS}, 2016.

\bibitem{huang2007labeled}
G.~B. Huang, M.~Ramesh, T.~Berg, and E.~Learned-Miller.
\newblock Labeled faces in the wild: A database for studying face recognition
  in unconstrained environments.
\newblock Technical report, Technical Report 07-49, University of
  Massachusetts, Amherst, 2007.

\bibitem{ilia2015face}
P.~Ilia, I.~Polakis, E.~Athanasopoulos, F.~Maggi, and S.~Ioannidis.
\newblock Face/off: Preventing privacy leakage from photos in social networks.
\newblock In {\em CCS}, 2015.

\bibitem{independent2012group}
{Independent JPEG Group}.
\newblock Independent {JPEG} group.
\newblock \url{http://www.ijg.org/}, 2012.

\bibitem{ioffe2015batch}
S.~Ioffe and C.~Szegedy.
\newblock Batch normalization: Accelerating deep network training by reducing
  internal covariate shift.
\newblock {\em arXiv:1502.03167}, 2015.

\bibitem{jarrett2009best}
K.~Jarrett, K.~Kavukcuoglu, Y.~Lecun, et~al.
\newblock What is the best multi-stage architecture for object recognition?
\newblock In {\em 2009 IEEE 12th International Conference on Computer Vision},
  pages 2146--2153. IEEE, 2009.

\bibitem{korayem2016enhancing}
M.~Korayem, R.~Templeman, D.~Chen, and D.~C.~A. Kapadia.
\newblock Enhancing lifelogging privacy by detecting screens.
\newblock In {\em Proceedings of the 2016 CHI Conference on Human Factors in
  Computing Systems}, 2016.

\bibitem{cifar}
A.~Krizhevsky and G.~Hinton.
\newblock Learning multiple layers of features from tiny images, 2009.

\bibitem{krizhevsky2012imagenet}
A.~Krizhevsky, I.~Sutskever, and G.~E. Hinton.
\newblock {ImageNet} classification with deep convolutional neural networks.
\newblock In {\em Advances in neural information processing systems}, 2012.

\bibitem{lander2001evaluating}
K.~Lander, V.~Bruce, and H.~Hill.
\newblock Evaluating the effectiveness of pixelation and blurring on masking
  the identity of familiar faces.
\newblock {\em Applied Cognitive Psychology}, 2001.

\bibitem{lecun2015deep}
Y.~LeCun, Y.~Bengio, and G.~Hinton.
\newblock Deep learning.
\newblock {\em Nature}, 2015.

\bibitem{mnist}
Y.~LeCun, C.~Cortes, and C.~J. Burges.
\newblock The mnist database of handwritten digits, 1998.

\bibitem{warondrugs-pic}
D.~Levene.
\newblock A police raid in {London} in 2011.
\newblock {\em The Guardian}, 2011.

\bibitem{melicher2016fast}
W.~Melicher, B.~Ur, S.~M. Segreti, S.~Komanduri, L.~Bauer, N.~Christin, and
  L.~F. Cranor.
\newblock Fast, lean and accurate: Modeling password guessability using neural
  networks.
\newblock In {\em Proceedings of USENIX Security}, 2016.

\bibitem{neustaedter2006blur}
C.~Neustaedter, S.~Greenberg, and M.~Boyle.
\newblock Blur filtration fails to preserve privacy for home-based video
  conferencing.
\newblock {\em ACM Transactions on Computer-Human Interaction (TOCHI)}, 2006.

\bibitem{newton2005preserving}
E.~M. Newton, L.~Sweeney, and B.~Malin.
\newblock Preserving privacy by de-identifying face images.
\newblock {\em IEEE transactions on Knowledge and Data Engineering}, 2005.

\bibitem{facescrub}
H.-W. Ng and S.~Winkler.
\newblock A data-driven approach to cleaning large face datasets.
\newblock In {\em IEEE International Conference on Image Processing (ICIP)},
  2014.

\bibitem{oh2016faceless}
S.~J. Oh, R.~Benenson, M.~Fritz, and B.~Schiele.
\newblock Faceless person recognition; privacy implications in social media.
\newblock {\em arXiv:1607.08438}, 2016.

\bibitem{oliphant2007python}
T.~E. Oliphant.
\newblock Python for scientific computing.
\newblock {\em Computing in Science \& Engineering}, 2007.

\bibitem{girlalone-pic}
U.~N.~O. on~Drugs and Crime.
\newblock India: Community vigilance rescues {Roshni}.
\newblock 2010.

\bibitem{parkhi2015deep}
O.~M. Parkhi, A.~Vedaldi, and A.~Zisserman.
\newblock Deep face recognition.
\newblock In {\em British Machine Vision Conference}, 2015.

\bibitem{pascanu2015malware}
R.~Pascanu, J.~W. Stokes, H.~Sanossian, M.~Marinescu, and A.~Thomas.
\newblock Malware classification with recurrent networks.
\newblock In {\em ICASSP}, 2015.

\bibitem{punnappurath2015face}
A.~Punnappurath, A.~N. Rajagopalan, S.~Taheri, R.~Chellappa, and
  G.~Seetharaman.
\newblock Face recognition across non-uniform motion blur, illumination, and
  pose.
\newblock {\em IEEE Transactions on Image Processing}, 2015.

\bibitem{p3}
M.-R. Ra, R.~Govindan, and A.~Ortega.
\newblock P3: Toward privacy-preserving photo sharing.
\newblock In {\em NSDI}, 2013.

\bibitem{rumelhart1988learning}
D.~E. Rumelhart, G.~E. Hinton, and R.~J. Williams.
\newblock Learning representations by back-propagating errors.
\newblock {\em Cognitive modeling}, 5(3):1, 1988.

\bibitem{ryan1998intrusion}
J.~Ryan, M.-J. Lin, and R.~Miikkulainen.
\newblock Intrusion detection with neural networks.
\newblock {\em Advances in neural information processing systems}, 1998.

\bibitem{schroff2015facenet}
F.~Schroff, D.~Kalenichenko, and J.~Philbin.
\newblock Facenet: A unified embedding for face recognition and clustering.
\newblock In {\em Proceedings of the IEEE Conference on Computer Vision and
  Pattern Recognition}, 2015.

\bibitem{simonyan2014very}
K.~Simonyan and A.~Zisserman.
\newblock Very deep convolutional networks for large-scale image recognition.
\newblock {\em arXiv preprint arXiv:1409.1556}, 2014.

\bibitem{sivakorn2016robot}
S.~Sivakorn, I.~Polakis, and A.~D. Keromytis.
\newblock I am robot:(deep) learning to break semantic image captchas.
\newblock In {\em 2016 IEEE European Symposium on Security and Privacy
  (EuroS\&P)}, 2016.

\bibitem{srivastava2014dropout}
N.~Srivastava, G.~E. Hinton, A.~Krizhevsky, I.~Sutskever, and R.~Salakhutdinov.
\newblock Dropout: a simple way to prevent neural networks from overfitting.
\newblock {\em JMLR}, 2014.

\bibitem{jpeg}
{Standard, JPEG}.
\newblock Information technology-digital compression and coding of
  continuous-tone still images-requirements and guidelines.
\newblock {\em International Telecommunication Union. CCITT recommendation},
  1992.

\bibitem{tran2016privacy}
L.~Tran, D.~Kong, H.~Jin, and J.~Liu.
\newblock Privacy-cnh: A framework to detect photo privacy with convolutional
  neural network using hierarchical features.
\newblock {\em AAAI 2016}, 2016.

\bibitem{venkatraman2014}
D.~Venkatraman.
\newblock Why blurring sensitive information is a bad idea.
\newblock \url{https://dheera.net/projects/blur}, 2014.

\bibitem{weyand2016planet}
T.~Weyand, I.~Kostrikov, and J.~Philbin.
\newblock Planet-photo geolocation with convolutional neural networks.
\newblock {\em arXiv:1602.05314}, 2016.

\bibitem{wilber2016can}
M.~J. Wilber, V.~Shmatikov, and S.~Belongie.
\newblock Can we still avoid automatic face detection?
\newblock In {\em WACV}, 2016.

\bibitem{youtubeface}
{YouTube Official Blog}.
\newblock Face blurring: when footage requires anonymity.
\newblock
  \url{https://youtube.googleblog.com/2012/07/face-blurring-when-footage-requires.html},
  2012.

\bibitem{youtubeblur}
{YouTube Official Blog}.
\newblock Face blurring: when footage requires anonymity.
\newblock
  \url{https://youtube-creators.googleblog.com/2016/02/blur-moving-objects-in-your-video-with.html},
  2016.

\bibitem{yuan2014droid}
Z.~Yuan, Y.~Lu, Z.~Wang, and Y.~Xue.
\newblock Droid-sec: Deep learning in {Android} malware detection.
\newblock In {\em ACM SIGCOMM Computer Communication Review}, 2014.

\bibitem{zagoruyko2015}
S.~Zagoruyko.
\newblock 92.45\% on {CIFAR-10 in Torch}.
\newblock \url{http://torch.ch/blog/2015/07/30/cifar.html}, 2015.

\bibitem{zhang2015beyond}
N.~Zhang, M.~Paluri, Y.~Taigman, R.~Fergus, and L.~Bourdev.
\newblock Beyond frontal faces: Improving person recognition using multiple
  cues.
\newblock In {\em CVPR}, 2015.

\bibitem{zhou2015naive}
E.~Zhou, Z.~Cao, and Q.~Yin.
\newblock Naive-deep face recognition: Touching the limit of {LFW} benchmark or
  not?
\newblock {\em arXiv:1501.04690}, 2015.

\end{thebibliography}

\begin{appendices}
\small
\section{Neural Network Architectures}

\subsection{MNIST Neural Network}
\label{app:mnist}
\begin{verbatim}
nn.Sequential {
  [input -> (1) -> (2) -> ... -> (11) -> (12) -> output]
  (1): nn.SpatialConvolutionMM(1 -> 32, 5x5)
  (2): nn.LeakyReLU(0.01)
  (3): nn.SpatialMaxPooling(3x3, 3,3)
  (4): nn.SpatialConvolutionMM(32 -> 64, 5x5)
  (5): nn.LeakyReLU(0.01)
  (6): nn.SpatialMaxPooling(2x2, 2,2)
  (7): nn.Reshape(256)
  (8): nn.Linear(256 -> 200)
  (9): nn.LeakyReLU(0.01)
  (10): nn.Dropout(0.500000)
  (11): nn.Linear(200 -> 10)
  (12): nn.LogSoftMax
}
\end{verbatim}

\subsection{CIFAR Neural Network}
\label{app:cifar}
\begin{verbatim}
nn.Sequential {
 [input -> (1) -> (2) -> (3) -> output]
 (1): nn.BatchFlip
 (2): nn.Copy
 (3): nn.Sequential {
   [input -> (1) -> (2) -> ... -> (53) -> (54) -> output]
   (1): nn.SpatialConvolution(3 -> 64, 3x3, 1,1, 1,1)
   (2): nn.SpatialBatchNormalization
   (3): nn.ReLU
  (4): nn.Dropout(0.300000)
   (5): nn.SpatialConvolution(64 -> 64, 3x3, 1,1, 1,1)
   (6): nn.SpatialBatchNormalization
   (7): nn.ReLU
   (8): nn.SpatialMaxPooling(2x2, 2,2)
   (9): nn.SpatialConvolution(64 -> 128, 3x3, 1,1, 1,1)
   (10): nn.SpatialBatchNormalization
   (11): nn.ReLU
   (12): nn.Dropout(0.400000)
   (13): nn.SpatialConvolution(128 -> 128, 3x3, 1,1, 1,1)
   (14): nn.SpatialBatchNormalization
   (15): nn.ReLU
   (16): nn.SpatialMaxPooling(2x2, 2,2)
   (17): nn.SpatialConvolution(128 -> 256, 3x3, 1,1, 1,1)
   (18): nn.SpatialBatchNormalization
   (19): nn.ReLU
   (20): nn.Dropout(0.400000)
   (21): nn.SpatialConvolution(256 -> 256, 3x3, 1,1, 1,1)
   (22): nn.SpatialBatchNormalization
   (23): nn.ReLU
   (24): nn.Dropout(0.400000)
   (25): nn.SpatialConvolution(256 -> 256, 3x3, 1,1, 1,1)
   (26): nn.SpatialBatchNormalization
   (27): nn.ReLU
   (28): nn.SpatialMaxPooling(2x2, 2,2)
   (29): nn.SpatialConvolution(256 -> 512, 3x3, 1,1, 1,1)
   (30): nn.SpatialBatchNormalization
   (31): nn.ReLU
   (32): nn.Dropout(0.400000)
   (33): nn.SpatialConvolution(512 -> 512, 3x3, 1,1, 1,1)
   (34): nn.SpatialBatchNormalization
   (35): nn.ReLU
   (36): nn.Dropout(0.400000)
   (37): nn.SpatialConvolution(512 -> 512, 3x3, 1,1, 1,1)
   (38): nn.SpatialBatchNormalization
   (39): nn.ReLU
   (40): nn.SpatialMaxPooling(2x2, 2,2)
   (41): nn.SpatialConvolution(512 -> 512, 3x3, 1,1, 1,1)
   (42): nn.SpatialBatchNormalization
   (43): nn.ReLU
   (44): nn.Dropout(0.400000)
   (45): nn.SpatialConvolution(512 -> 512, 3x3, 1,1, 1,1)
   (46): nn.SpatialBatchNormalization
   (47): nn.ReLU
   (48): nn.Dropout(0.400000)
   (49): nn.SpatialConvolution(512 -> 512, 3x3, 1,1, 1,1)
   (50): nn.SpatialBatchNormalization
   (51): nn.ReLU
   (52): nn.SpatialMaxPooling(2x2, 2,2)
   (53): nn.View(512)
   (54): nn.Sequential {
     [input -> (1) -> (2) -> ... -> (5) -> (6) -> output]
     (1): nn.Dropout(0.500000)
     (2): nn.Linear(512 -> 512)
     (3): nn.BatchNormalization
     (4): nn.ReLU
     (5): nn.Dropout(0.500000)
     (6): nn.Linear(512 -> 10)
   }
 }
}
\end{verbatim}

\subsection{AT\&T Neural Network}
\label{app:att}
\begin{verbatim}
nn.Sequential {
  [input -> (1) -> (2) -> ... -> (14) -> (15) -> output]
  (1): nn.SpatialConvolutionMM(1 -> 32, 3x3, 1,1, 1,1)
  (2): nn.LeakyReLU(0.01)
  (3): nn.SpatialMaxPooling(2x2, 2,2)
  (4): nn.SpatialConvolutionMM(32 -> 64, 3x3, 1,1, 1,1)
  (5): nn.LeakyReLU(0.01)
  (6): nn.SpatialMaxPooling(2x2, 2,2)
  (7): nn.SpatialConvolutionMM(64 -> 128, 3x3, 1,1, 1,1)
  (8): nn.LeakyReLU(0.01)
  (9): nn.SpatialMaxPooling(3x3, 3,3)
  (10): nn.Reshape(8064)
  (11): nn.Linear(8064 -> 1024)
  (12): nn.LeakyReLU(0.01)
  (13): nn.Dropout(0.500000)
  (14): nn.Linear(1024 -> 40)
  (15): nn.LogSoftMax
}
\end{verbatim}

\subsection{FaceScrub Neural Network}
\label{app:facescrub}
\begin{verbatim}
FaceScrub	
nn.Sequential {
  [input -> (1) -> (2) -> ... -> (13) -> (14) -> output]
  (1): nn.SpatialConvolutionMM(1 -> 32, 3x3, 1,1, 1,1)
  (2): nn.LeakyReLU(0.01)
  (3): nn.SpatialMaxPooling(2x2, 2,2)
  (4): nn.SpatialConvolutionMM(32 -> 64, 3x3, 1,1, 1,1)
  (5): nn.LeakyReLU(0.01)
  (6): nn.SpatialMaxPooling(2x2, 2,2)
  (7): nn.SpatialConvolutionMM(64 -> 128, 3x3, 1,1, 1,1)
  (8): nn.LeakyReLU(0.01)
  (9): nn.SpatialMaxPooling(2x2, 2,2)
  (10): nn.Reshape(100352)
  (11): nn.Linear(100352 -> 1024)
  (12): nn.LeakyReLU(0.01)
  (13): nn.Dropout(0.500000)
  (14): nn.Linear(1024 -> 530)
}
\end{verbatim}

\normalsize

\end{appendices}

\end{document}